\documentclass[useAMS,usenatbib]{mn2e}
\usepackage{graphicx}

\title[Bolometric luminosities of type Ia SN progenitors]{Upper limits on bolometric luminosities of three type Ia supernova progenitors - New results in the ongoing \textit{Chandra} archival search for type Ia supernova progenitors}
\author[M.T.B. Nielsen, R. Voss and G. Nelemans]{M.T.B. Nielsen$^{1}$\thanks{E-mail:
m.nielsen@astro.ru.nl}, R. Voss$^{1}$ and G. Nelemans$^{1,2}$\\
$^{1}$Department of Astrophysics, IMAPP, Radboud University Nijmegen, PO Box 9010, NL-6500 GL Nijmegen, the Netherlands\\
$^{2}$Institute for Astronomy, KU Leuven, Celestijnenlaan 200D, 3001 Leuven, Belgium}
\begin{document}

\date{Accepted -. Received \today; in original form -}

\pagerange{\pageref{firstpage}--\pageref{lastpage}} \pubyear{2011}

\maketitle

\label{firstpage}

\begin{abstract}
We present analysis of \textit{Chandra} archival, pre-explosion data of the positions of three nearby ($<$ 25 Mpc) type Ia supernovae, SN2011iv, SN2012cu \& SN2012fr. No sources corresponding to the progenitors were found in any of the observations. Combining all sources with well defined backgrounds does not reveal any evidence for X-ray emission from the progenitors either. We calculated upper limits on the bolometric luminosities of the progenitors, under the assumption that they were black bodies with effective temperatures between 30 and 150 eV, corresponding to 'canonical' supersoft X-ray sources. The upper limits of SN2012fr straddles the Eddington luminosity of canonical supersoft sources, but fainter canonical supersoft sources cannot be ruled out by this study. We also compare our upper limits with known compact binary supersoft X-ray sources. This study is a continuation of the campaign to directly detect or constrain the X-ray characteristics of pre-explosion observations of nearby type Ia supernova progenitors; with the results reported in Nielsen, Voss \& Nelemans (see reference in Introduction), the number of nearby type Ia supernovae for which pre-explosion images are available in the \textit{Chandra} archive is now 13 and counting.
\end{abstract}


\section{Introduction} \label{Sect:Introduction}
Type Ia supernovae (SNe) are of crucial imporance for both cosmology and galactic chemical evolution. However, the nature of the progenitor systems giving rise to type Ia SNe remains unresolved. Two scenarios are normally considered: the single-degenerate (SD), in which a carbon-oxygen white dwarf (WD) grows in mass by accreting hydrogen-rich material from a companion \citep{Whelan.Iben.1973}, and the double-degenerate (DD) scenario, where a double WD binary system slowly spirals in and finally merges \citep{Webbink.1984,Iben.Tutukov.1984}. Beyond these two major scenarios, there are also a number of alternative scenarios, such as the 'core degenerate' scenario \citep{Kashi.Soker.2011}. In all scenarios, the end result is a carbon-oxygen WD with a mass at or above a critical mass ($\sim 1.38 \mathrm{ M}_{\odot}$), where the density and temperature of the WD interior is high enough to facilitate a runaway of unstable thermonuclear burning, which completely unbinds the WD in a supernova explosion. The process produces radioactive iron-group elements, and the subsequent decay of these powers a characteristic light-curve that can be used as standardizable cosmological candles \citep{Philips.1993}, e.g. to measure the accelerating expansion of the Universe \citep{Riess.et.al.1998,Perlmutter.et.al.1999}.

In the SD scenario, the accreting WD is expected to be a supersoft X-ray source (SSS), with a black body temperature between 30 and 150 eV, and luminosity of $\sim 10^{37}-10^{38} \mathrm{erg/s}$ \citep{van.den.Heuvel.et.al.1992,Kahabka.van.den.Heuvel.1997}. For sources closer than $\sim$ 25 Mpc, such sources should be detectable with the \textit{Chandra} X-ray observatory. With the aim of directly detecting a progenitor of a type Ia SN immediately prior to the explosion, a search of archival pre-explosion \textit{Chandra} images was undertaken by \citet{Voss.Nelemans.2008}, resulting in the possible, but ambiguous \citep{Roelofs.et.al.2008} detection of the X-ray emission of the progenitor of SN2007on. Since then, a systematic search has been conducted, resulting in upper limits of the bolometric luminosities of ten additional nearby type Ia SNe, SN2002cv, SN2003cg, SN2004W, SN2006X, SN2006dd, SN2006mr, SN2007gi, SN2007sr, SN2008fp, and SN2011fe, presented in \citet{Nielsen.et.al.2012}.

In this article, we expand on the results of \citet{Nielsen.et.al.2012} by presenting upper limits on the bolometric luminosities of three additional supernovae: SN2011iv (detected by Drescher (CBET 2940), classified by Chen \& Wang and Stritzinger (CBET 2940)), SN2012cu (detected by Itagaki (CBET 3146), classified by Marion \& Milisavljevic (CBET 3146)), and SN2012fr (detected by Klotz and the \textit{TAROT} collaboration (CBET 3275) and classified by Childress et al. (CBET 3275)). Based on the \textit{Chandra} observations, we calculate upper limits on the bolometric luminosities of the progenitors, assuming black-body spectra for four effective temperatures in the super-soft range ($kT_{\mathrm{BB}}=$ 30, 50, 100, and 150 eV). The method used is identical to the one used in \citet{Nielsen.et.al.2012}, so the results are immediately comparable, and together the 13 upper limits, plus the ambiguous case of SN2007on (see \citealt{Voss.Nelemans.2008,Roelofs.et.al.2008}), constitute the complete set of currently known nearby ($<$ 25 Mpc) type Ia SNe with pre-explosion observations in the \textit{Chandra} archive. We compare the calculated upper limits with the luminosities of known SSS in nearby galaxies.

We note that the current ambiguity concerning the possible direct detection of the progenitor of SN2007on is unlikely to be resolved until new \textit{Chandra} observations of the position of the SN are available. Because of this, we do not include SN2007on in our sample of nearby type Ia SNe with pre-explosion \textit{Chandra} observations.

In Section \ref{Sect:Observations} the \textit{Chandra} observations used in this study will be described. Section \ref{Sect:Data.Reduction} explains the methods employed in the data analysis of the observations, and gives the results. Section \ref{Sect:Discussion} discusses our results, and Section \ref{Sect:Conclusion} concludes.

\section{Observations} \label{Sect:Observations}
We searched the \textit{Chandra} Data Archive and found pre-explosion observations with the Advanced CCD Imaging Spectrometer (ACIS-S) at the positions of three nearby ($>$25 Mpc) type Ia SNe: SN2011iv, SN2012cu, and SN2012fr. No obvious sources corresponding to X-ray emitting progenitors were found on any of the pre-explosion images.

For SN2012cu, only a single pre-explosion observation exists, while several epochs of pre-explosion observations are available for SN2012cu and SN2012fr. For SN2012fr two additional, pre-explosion observations (obs IDs 13920 \& 13921), both of them relatively deep (90 and 110 kiloseconds, respectively), will become available in april 2013. Our analyses do not take these two, currently proprietary observation epochs into account.

The observations used in this study are listed in Table \ref{Table:Observations}.

\begin{table*}
 \begin{minipage}{140mm}
 \caption{\textit{Chandra} observations used in this study. All observations are with the ACIS-S detector.}
 \centering
  \begin{tabular}{@{}c c c c c c @{}}
  \hline
  \textit{Chandra}	& exposure	& pointing  			& SN 		& SN			& observation \\
      observation	& time		& (RA, DEC)			&		& explosion		& date \\
			& [ks]		&				& 		& date			& \\
  \hline
  \hline
    2942		& 29.24		& (03:38:52.00,	-35:35:34.00)	& 2011iv	& 2011-12-02 		& 2003-02-13 \\
    4174		& 45.67		& (03:38:49.58,	-35:34:36.34)	& 2011iv	& 2011-12-02		& 2003-05-28 \\
    9798		& 18.3		& (03:38:51.00,	-35:34:31.00)	& 2011iv	& 2011-12-02		& 2007-12-24 \\
    9799		& 21.29		& (03:38:51.00,	-35:34:31.00)	& 2011iv	& 2011-12-02		& 2007-12-27 \\
    3999		& 4.71		& (12:53:29.20	+02:10:06.20)	& 2012cu	& 2012-06-14		& 2003-02-14 \\
    3554		& 14.61		& (03:33:36.40	-36:08:25.00)	& 2012fr	& 2012-10-27		& 2002-12-24 \\
    6868		& 14.61		& (03:33:36.40	-36:08:25.00)	& 2012fr	& 2012-10-27		& 2006-04-17 \\
    6869		& 15.54		& (03:33:36.40	-36:08:25.00)	& 2012fr	& 2012-10-27		& 2006-04-20 \\
    6870		& 14.55		& (03:33:36.40	-36:08:25.00)	& 2012fr	& 2012-10-27		& 2006-04-23 \\
    6871		& 13.36		& (03:33:36.40	-36:08:25.00)	& 2012fr	& 2012-10-27		& 2006-04-10 \\
    6872		& 14.62		& (03:33:36.40	-36:08:25.00)	& 2012fr	& 2012-10-27		& 2006-04-12 \\
    6873		& 14.64		& (03:33:36.40	-36:08:25.00)	& 2012fr	& 2012-10-27		& 2006-04-14 \\
    13920		& 88.53		& (03:33:36.40 	-36:08:25.00)	& 2012fr	& 2012-10-27		& 2012-04-09 \\
    13921		& 108.2		& (03:33:36.40 	-36:08:25.00)	& 2012fr	& 2012-10-27		& 2012-04-12 \\
  \hline
\end{tabular} \label{Table:Observations}
\end{minipage}
\end{table*}

\section{Data Reduction \& Results} \label{Sect:Data.Reduction}
The observations mentioned above were analyzed with the \textsc{ciao} 4.3 software suite, and followed the same procedure as the one described in \citet{Nielsen.et.al.2012}. For ease of reference, all tables have been kept in the same format as that used in \citet{Nielsen.et.al.2012}.

All images were checked for photons of any energy at the positions of the SNe. No indications of the presence of sources were found. The images were then filtered to only include photons between 300 and 1000 eV, since below approximately 300 eV, the \textit{Chandra} ACIS detector is unreliable, and above 1 keV, we expect no photons from the sources.

As our data model, we assumed an absorbed black-body and used the spectral models {\tt xsphabs} and {\tt xsbbody} (corresponding to {\tt phabs} and {\tt bbody} in XSPEC). Spectral weights files were generated for four different black-body temperatures, $kT_{\mathrm{BB}}=$ 30, 50, 100 \& 150 eV, taking into account the absorbing columns mentioned below. We then generated exposure maps from the spectral weights files. Multiple, pre-explosion epochs of observations exits for SN2011iv and SN2012fr, and for these SNe we combined the binned images to get deeper observations.

The distances to the SNe were taken to be identical to those of their host galaxies, as found in the NED online database\footnote{http://ned.ipac.caltech.edu/}

For SN2012fr, \citet{Schlafly.Finkbeiner.2011} give the V-band absorption (in magnitudes) in the direction of the SN as $A_V=0.056$ (see also ATEL\#4535), and from this the neutral hydrogen column can be calculated, using the relation $N_H = 2.21\cdot10^{21} A_V$, where $N_H$ is the neutral hydrogen column in $\mathrm{cm}^{-2}$ \citep{Guver.Ozel.2009}. For SN2011iv, \citet{Foley.et.al.2012} found negligible extinction in the host galaxy, and for SN2012cu, no values for the hydrogen column, reddening or extinction could be found in the literature. Therefore, for SN2011iv and SN2012cu we used the value for the Galactic column found in \citet{Dickey.Lockman.1990}, which is referenced with CIAO's COLDEN tool. The host galaxies, distances, and columns for the SNe analysed in this study are summarised in Table \ref{Table:Host.Gals.Dist.Columns}.

\begin{table*}
 \begin{minipage}{140mm} 
\caption{Host galaxies, distances and total hydrogen columns for each of the SNe analyzed in this study.}
 \centering
  \begin{tabular}{@{}c c c c c c @{}}
  \hline
  supernova 	& host		& distance	& absorbing		& reference	 			\\
		& galaxy	& [Mpc]		& column		& for column				\\
		&		& (from NED)	& [$N_H$ cm$^{-2}$]	& values				\\
  \hline
  \hline
  2011iv 	& NGC 1404 	& 25.0		& 1.51$\cdot10^{20}$	& \citet{Dickey.Lockman.1990} 		\\
  \hline
  2012cu 	& NGC 4772 	& 13.3		& 1.72$\cdot10^{20}$	& \citet{Dickey.Lockman.1990} 		\\
  \hline
  2012fr	& NGC 1365	& 20.7		& 1.24$\cdot10^{20}$	& \citet{Schlafly.Finkbeiner.2011}	\\
  \hline
\end{tabular} \label{Table:Host.Gals.Dist.Columns}
\end{minipage}
\end{table*}

For each observation, we used a circular aperture of 4.5 pixel radius, which covers more than 90\% of the point-spread function of a theoretical point source. This aperture contains the background plus a Poissonian realization of the expected number of photons from a source. For this photon count, $N_{\mathrm{obs}}$, we found the maximum average number of counts $\mu$, for which the probability $P$ of observing $N_{\mathrm{obs}}$ photons is within 3$\sigma$, under the assumption of Poissonian statistics, see e.g. \citet{Gehrels.1986}: $P \left( \mu, N \leq N_{\mathrm{obs}} \right) \leq 0.0013$. This $\mu$ represents the 3$\sigma$ upper limit of any progenitor including background. We found the upper limit to the luminosity of the source according to the formula,

\begin{eqnarray}
 L_{X,UL} = 4 \pi \frac{\left( \mu - b \right) \langle E_{\gamma} \rangle d^2}{\zeta} \label{Eq:L_X}
\end{eqnarray}
where $b$ is the expected background for a circular aperture of radius 4.5 pixels, $\langle E_{\gamma} \rangle$ is the average energy of the photons found from the absorbed XSPEC model for the assumed spectrum, $d$ is the distance to the SN and $\zeta$ is the value of the exposure map for the given spectrum at the position of the SN on the detector.

We then corrected the calculated luminosities for interstellar absorption, using the columns listed in Table \ref{Table:Host.Gals.Dist.Columns}, to obtain the unabsorbed luminosities of the sources. As a last step, we scaled the calculated supersoft X-ray luminosities to obtain bolometric luminosities.

Table \ref{Table:Upper.Limits} lists the results of our analysis. The individual \textit{Chandra} images are shown on Figures \ref{Fig:SN2011iv}-\ref{Fig:SN2012fr}. For SN2011iv and SN2012fr, these images are the combined images for all the available observation epochs, while Figure \ref{Fig:SN2012cu} shows the single pre-explosion \textit{Chandra} observation in existence of the position of SN2012cu.

\begin{table*}
 \begin{minipage}{.85\textwidth} 
  \caption{Nearby ($<$ 25 Mpc) SNe Ia with pre-explosion images, upper limit bolometric luminosities.}
  \centering
  \begin{tabular}{@{} c c c c c c c c @{}}
  \hline
   Supernova	& SN position	 	& pre-explosion		& exposure	& total		& 3-$\sigma$ counts & value of		& unabsorbed 3-$\sigma$ \\
		&[RA, DEC]		& \textit{Chandra}	& time		& counts	& in		& exposure		& upper limit \\
		&			& observations		& [ks]		& in		& source	& map at		& bolometric \\
		&			&			&		& source	& region	& position		& luminosity\footnote{for 30 eV, 50 eV, 100 eV, \& 150 eV, respectively.}\\
		&			&			&		& region\footnote{for combined image if several observation epochs exits} & region	& [s$\cdot$cm$^{2}$]	& [erg/s] \\
 \hline
 \hline
 2011iv	& (03:38:51.35, 		&2942, 4174, 9798, 9799& 114.5		& 243		& 289.8		& 3.65$\cdot10^{6}$	& 1.22$\cdot10^{41}$\\
	&  -35:35:32.0)			& 			& 		& 		& 		& 5.51$\cdot10^{6}$	& 5.55$\cdot10^{39}$\\
	& 				& 			& 		& 		& 		& 1.31$\cdot10^{7}$	& 6.07$\cdot10^{38}$\\
	& 				& 			& 		& 		& 		& 2.06$\cdot10^{7}$	& 3.56$\cdot10^{38}$\\
\hline
 2012cu	& (12:53:29.35,			& 3999			& 4.71		& 1		& 8.9		& 236537		& 6.89$\cdot10^{40}$\\
	&  +02:09:39.0)			& 			& 		& 		& 		& 327195		& 3.38$\cdot10^{39}$\\
	&				& 			& 		& 		& 		& 651910		& 4.35$\cdot10^{38}$\\
	&				& 			& 		& 		& 		& 941337		& 2.77$\cdot10^{38}$\\
\hline
 2012fr	& (03:33:35.99,		 	&3554, 6868, 6869, 6870,& 298.66	& 5		& 16.03		& 7.41$\cdot10^{6}$	& 6.64$\cdot10^{39}$\\
	& -36:07:37.7)			& 6871, 6872, 6873, 13920&		& 		& 		& 1.13$\cdot10^{7}$	& 3.00$\cdot10^{38}$\\
	&				& 13921			& 		& 		& 		& 2.74$\cdot10^{7}$	& 3.27$\cdot10^{37}$\\
	&				& 			& 		& 		& 		& 4.32$\cdot10^{7}$	& 1.93$\cdot10^{37}$\\
\hline
\end{tabular} \label{Table:Upper.Limits}
\end{minipage}
\end{table*}

\section{Discussion} \label{Sect:Discussion}
This study is a continuation of that presented in \citet{Nielsen.et.al.2012}. It increases the total number of nearby ($<$ 25 Mpc) type Ia SNe with pre-explosion images in \textit{Chandra} to 13. None of these show any evidence for an X-ray progenitor. We note that for  SN2007on an X-ray source was detected, see \citet{Voss.Nelemans.2008}, however, whether this actually was a detection or a chance alignment remains undetermined, as explained in \citet{Roelofs.et.al.2008}. We therefore disregard SN2007on in the following.

\begin{figure}
 \centering
  \includegraphics[width=1.0\linewidth]{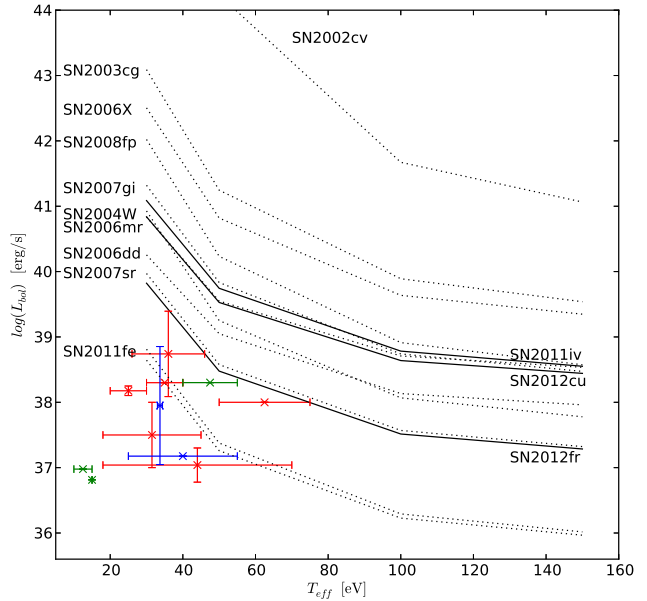} 
 \caption{Comparison between bolometric luminosity 3-$\sigma$ upper limits found in this paper and \citet{Nielsen.et.al.2012} with bolometric luminosities of known supersoft X-ray sources in nearby galaxies, as taken from \citet{Greiner.2000}. The dotted lines and SN designations on the left (center for SN2002cv) are for the 10 upper limits from the 2012 paper, while the 3 new upper limits presented in this paper are plotted with solid lines and SN designations on the right. The green, blue and red crosses are known compact binary and symbiotic SSSs in the Milky Way, SMC and LMC, respectively.}
   \label{Fig:Comparison.Upper.Limits.3sigma}
\end{figure}

In Figure \ref{Fig:Comparison.Upper.Limits.3sigma} we compare the upper limits found in this study and in \citet{Nielsen.et.al.2012} with the known SSSs of the Magellanic Clouds and the Milky Way. As can be seen, the upper limits of SN2012fr are comparable to those of SN2007sr, which narrowly straddles the parameter space where we expect to find canonical SSSs, i.e. $kT_{\mathrm{BB}} = 30-100$ eV \& $L_{\mathrm{bol}}\simeq 10^{37} - 10^{38}$ erg/s. Even for SN2012fr, we can rule out the brightest SSSs as progenitors, provided they are unobscured by local material. However, for all the SNe in the sample, a progenitor with a low effective temperature and/or bolometric luminosity would be permitted by the observations. SN2011fe remains the most constraining case by far, which primarily stems from the fact that it was the most nearby type Ia SN since the launch of \textit{Chandra}, and it took place in a well-observed galaxy (M101).

\begin{figure}
 \centering
  \includegraphics[width=1.0\linewidth]{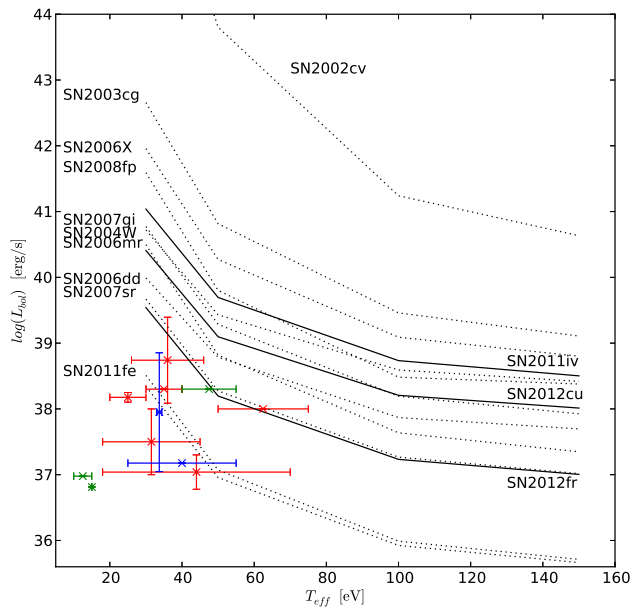} 
 \caption{1-$\sigma$ version of Figure \ref{Fig:Comparison.Upper.Limits.3sigma}.}
   \label{Fig:Comparison.Upper.Limits.1sigma}
\end{figure}

\begin{figure}
 \centering
  \includegraphics[width=1.0\linewidth]{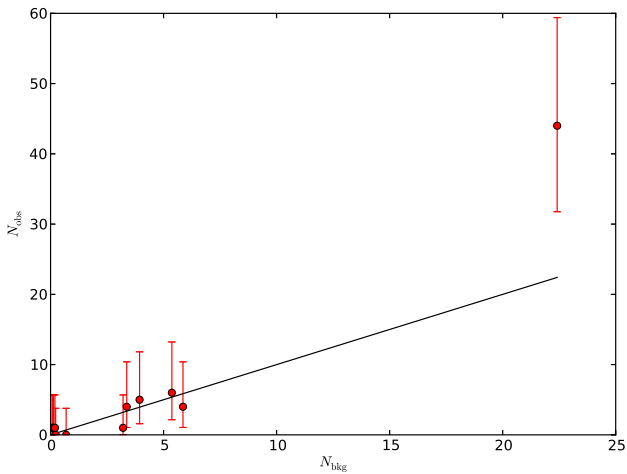} 
 \caption{Source counts ($N_{\mathrm{obs}}$) vs. expected background counts in the 4.5 pixel aperture source region ($N_{\mathrm{bkg}}$). Error bars correspond to 2-$\sigma$ confidence level. For ease of reference, the point corresponding to SN2011iv is not shown, as the count rates is very large (243) in comparison to the other points. The outlying point corresponds to SN2006mr.}
   \label{Fig:Source_vs_background_counts}
\end{figure}

Rather than just looking at the individual upper limits, we can use the fact that we now have a 13 consistently calculated upper limits to study the sample of objects. The first thing to note is that the comparison in Figure \ref{Fig:Comparison.Upper.Limits.3sigma} is in a sense inconsistent: we compare 3-$\sigma$ upper limits with data with 1-$\sigma$ error bars. For individual systems, that is probably the right thing to do, but as a sample, a comparison using 1-$\sigma$ upper limits may be more relevant. Therefore, we show the same plot with 1-$\sigma$ upper limits in Figure \ref{Fig:Comparison.Upper.Limits.1sigma}. The 1-$\sigma$ upper limits of SN2007sr and SN2012fr are probing the $kT_{\mathrm{BB}}$, $L_{\mathrm{bol}}$-parameter space of known SSSs. Furthermore, we can have a look in more detail at the individual observations and how they compare to the expected background as a sample. In Figure \ref{Fig:Source_vs_background_counts} we compare the observed counts in the source region for the SNe in our sample, except SN2006mr and SN2011iv, with the number of counts expected from the background in the same region. Evidently, the pre-explosion source counts for all the SNe on the figure are consistent with background only (i.e. no source) to within 2-$\sigma$ confidence; this is also the case for SN2011iv (not shown on the figure). We note that at this confidence level, the pre-explosion source count on the position of SN2006mr (also not shown) is formally a detection; however, for a sample of 13 systems, $\sim$ 4.2 of them are expected to be detections when considering 1-$\sigma$ limits, and we should even expect one 2-$\sigma$ detection as well. So, for purely statistical reasons, the fact that SN2006mr is a 2-$\sigma$ detection should not come as a big surprise. Furthermore, SN2006 is located on a very uneven background region close to the center of its host galaxy, and the exact number of background photons is difficult to determine without additional information concerning the X-ray background (see \citealt{Nielsen.et.al.2012} for details).

\begin{figure}
 \centering
  \includegraphics[width=1.0\linewidth]{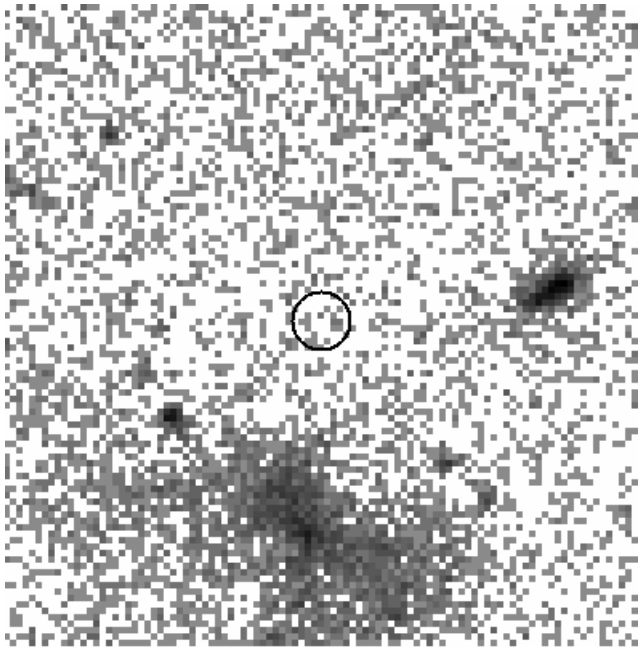}
 \caption{Stacked pre-explosion images of SN2002cv, SN2003cg, SN2004W, SN2006X, SN2006dd, SN2007gi, SN2007sr, SN2008fp, SN2011fe, SN2012cu, and SN2012fr. Pre-explosion images of SN2006mr and SN2011iv were excluded due to uneven backgrounds.}
   \label{Fig:Stacked_image_allSNe}
\end{figure}

\begin{figure}
 \hspace{0pt}\mbox{
 \begin{minipage}[c]{250pt}
  \centering
  \includegraphics[width=\textwidth]{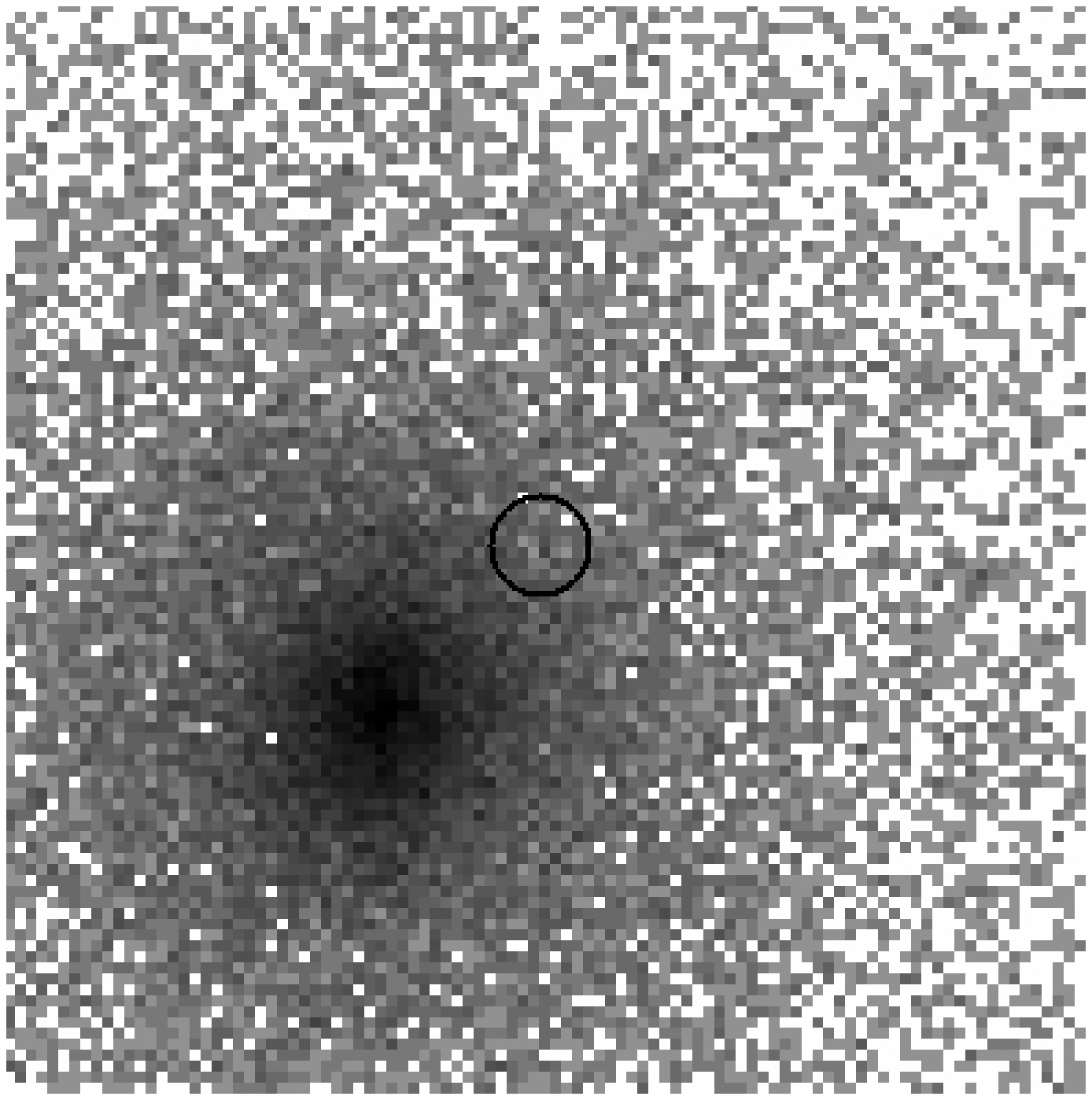}
  \caption{Part of \textit{Chandra} observations 2942, 4174, 9798 \& 9799. The circle corresponds to an aperture of 4.5 pixels at the position of SN2011iv.}\label{Fig:SN2011iv}
 \end{minipage}
 \hspace{.03\textwidth}
 \begin{minipage}[c]{250pt}
  \centering
  \includegraphics[width=\textwidth]{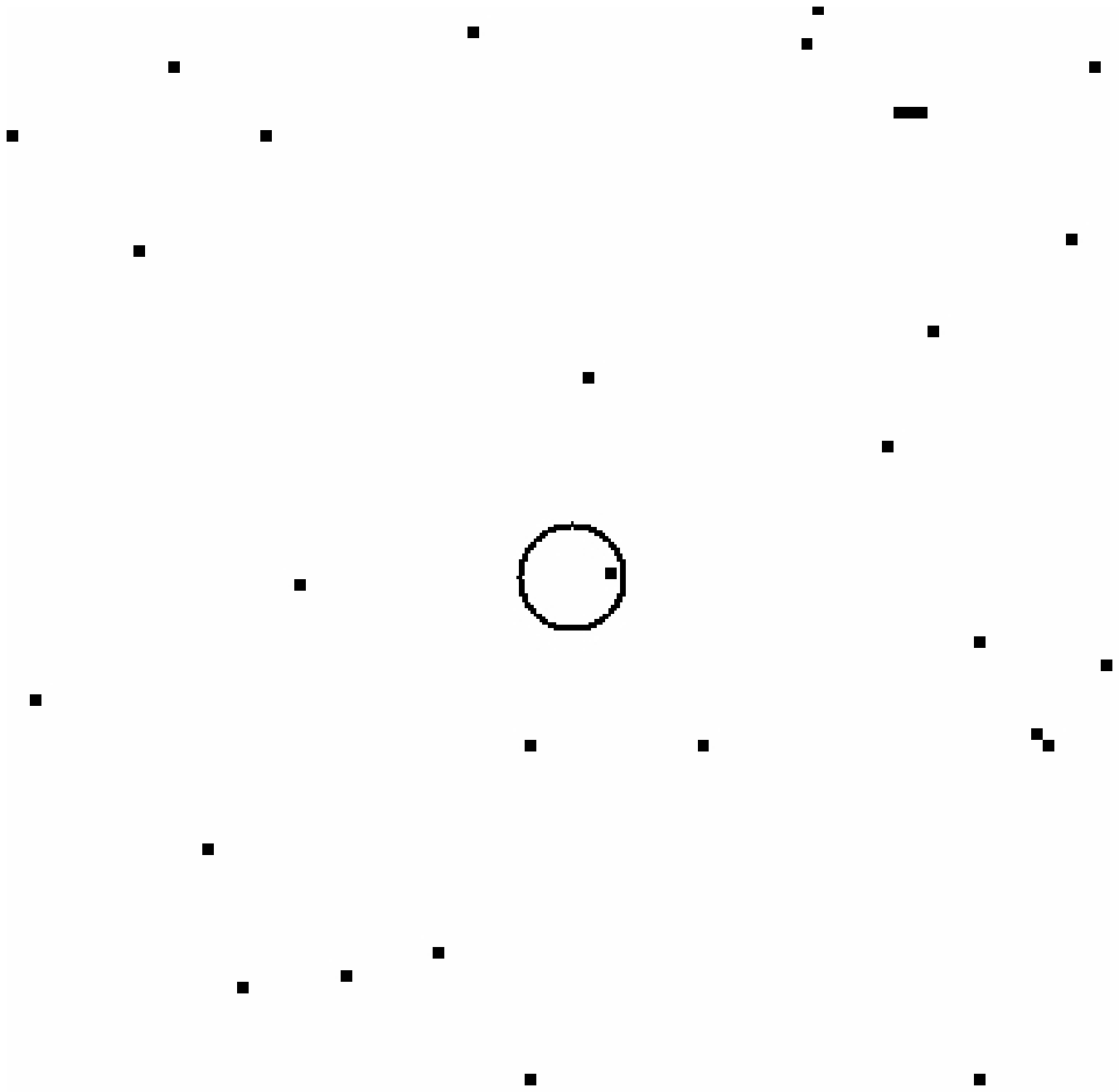}
  \caption{Part of combined image consisting of \textit{Chandra} observation 3999. The circle corresponds to an aperture of 4.5 pixels at the position of SN2012cu}\label{Fig:SN2012cu}
 \end{minipage}
}
 \hspace{0pt}\mbox{
 \begin{minipage}[c]{250pt}
  \centering
  \includegraphics[width=\textwidth]{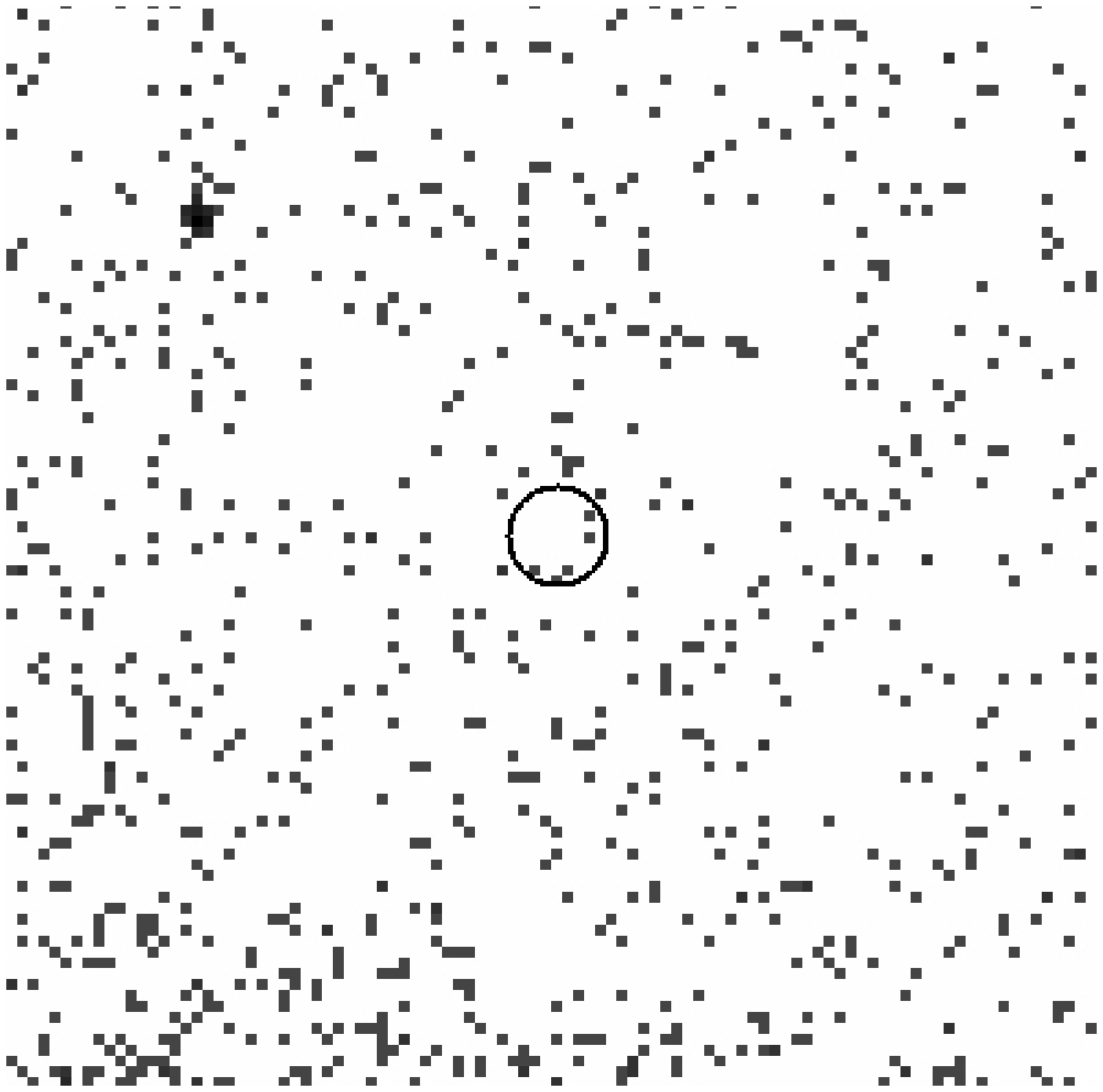}
  \caption{Part of \textit{Chandra} observations 3554, 6868, 6869, 6870, 6871, 6872, 6873, 13920, \& 13921. The circle corresponds to an aperture of 4.5 pixels at the position of SN2012fr.}\label{Fig:SN2012fr}
 \end{minipage}
 \hspace{.03\textwidth}
}
\begin{center}
\end{center}
\end{figure}

Figure \ref{Fig:Stacked_image_allSNe} shows a stacked image of the SNe that are located on a reasonably even background (i.e. excluding SN2006mr, S2006dd and SN2011iv). If there were a general tendency towards more photons in the source regions of each image it would become gradually clearer as the images are added. We find no evidence for such a stack-up of photons.

To make a more qualitative statement, we compared the number of observed pre-explosion photons for each SN position with two hypotheses: the null hypothesis (hypothesis $H_0$) that there is no source at the position of the SNe prior to the SN explosion. The alternative hypothesis (hypothesis $H_1$) is that there is a naked (i.e. unobscured by local phenomena), 'canonical' SSS ($L_{\mathrm{bol}}=10^{38}$ erg/s, $kT_{\mathrm{BB}}=50$ eV) at the positions of the SNe prior to the SN explosion. We calculated the expected number of counts for canonical SSSs at the distances given in Table \ref{Table:Host.Gals.Dist.Columns} and Table 1 in \citet{Nielsen.et.al.2012}. We rescaled this bolometric flux to the supersoft X-ray flux in \textit{Chandra}'s 0.3-1.0 keV band, then found the absorbed flux using the relevant absorbing column given in Table \ref{Table:Host.Gals.Dist.Columns} and Table 1 in \citet{Nielsen.et.al.2012}, and inserted this into Equation \ref{Eq:L_X} to find the expected number of photons from such a source. The number of photons expected if no source is present is simply the number of background photons. We then calculated the Poissonian probabilities $P$ of detecting the observed number of counts for the two hypotheses. The ratio $P_{H_1}/P_{H_0}$ is the likelihood ratio. Table \ref{Table:Compare_hypotheses} shows the likelihood ratios for the alternative and null hypotheses for the entirety of our nearby SN sample. In general, a likelihood ratio larger than 1 favours hypothesis 1, i.e. that there is a canonical SSS present at the position of the SN prior to the explosion. Conversely, a number smaller than 1 favours the null hypothesis that there is no source present. Values close to 1 indicate that we cannot discriminate between the two hypotheses.

The strongest conclusion we can make from the individual likelihood ratios in Table \ref{Table:Compare_hypotheses} is that for SN2011fe the presence of a naked, canonical SSS is strongly ruled out. For SN2007sr and SN2012fr, it appears that we can rule out a source, but not with very strong confidence. Conversely, for SN2008fp, SN2011iv, and SN2012cu, the presence of sources are favoured, but again, only weakly so. For SN2006mr, we seem to have a strong indication of the presence of a source, however, as Fig. 7 in \citet{Nielsen.et.al.2012} shows, the background of the position of the progenitor is quite uneven. Therefore, whether this is actually a detection of a source is unclear. We also note that for SN2002cv the probabilities for either hypotheses are the same. This is due to the large absorbing column for that SN, and it essentially means that including the pre-explosion observations for SN2002cv in our analysis confers no information as to whether there is a source present or not.

By taking the product of the likelihood ratios, i.e. $\Pi (P_{H_1}/P_{H_0})$, we can determine whether or not the sample as a whole gives us reason to believe that sources are actually present in the pre-explosion observations. Due to the extremely small likelihood ratio for SN2011fe, the combined likelihood ratio of the entire sample is tiny ($\sim 2.2 \cdot 10^{-12}$).  However, if we accept that we can rule out a naked, canonical SSS as the progenitor of SN2011fe and leave it out, along with the problematic case of SN2006mr, we get a combined likelihood ratio of $\sim$0.99. To properly interpret this number, we conduct a simulation to assess the probability that this combined likelihood ratio could have been observed by chance even if in reality there were canonical SSSs present in all osbervations. In order to do so, we assumed that a source was present and simulated 100,000 Poissonian realisations of the source counts for each pre-SN observation with expectation values as per a canonical SSS. These were multiplied into combined likelihood ratios for the subsample excluding SN2011fe and/or SN2006mr. We then counted the fraction of cases where the simualtions resulted in a likelihood ratio lower than the observed value. For the full sample, there is no chance to find the extremely low likelihood ratio by chance (0 instances out of 100,000). For the sub-sample that excludes SN2011fe and SN2006mr, we find a value of $\sim$0.99 or lower in only $\sim$ 14\% of the cases, i.e. the low value could be found by chance even if there were sources present in the pre-SN observations, but it is rather unlikely. This confirms our interpretation of the likelihood ratio, that it is a hint against canonical SSS progenitors being present for \emph{all} supernovae, but not a very strong hint.

We conclude that the sample as a whole strongly disfavours the hypothesis that there are sources present, but this is dominated by the strong upper limits of SN2011fe. If we accept SN2011fe as a non-detection and exclude it from our sample, it essentially becomes impossible to determine with a high level of confidence whether sources are present or not. Removing the problematic case of SN2006mr favours the non-presence of sources, to approximately the 1-$\sigma$ level.

\begin{table*}
 \begin{minipage}{.65\textwidth}
\caption{Comparison of hypotheses $H_0$ and $H_1$ (see text for details).}
 \centering
  \begin{tabular}{@{} c c c c c c c c c @{}}
  \hline
  SN		& counts	& photons		& photons		& $P_{H_1}$		& $P_{H_0}$		& $P_{H_1}/P_{H_0}$	\\
		& observed	& expected,		& expected,		& 			&			& 			\\
		&		& canonical		& no source		& 			&			& 			\\
		&		& SSS			& 			&			&			& 			\\
  \hline
  \hline
  2002cv	& 1		& 1.01$\cdot10^{-1}$	& 1.01$\cdot10^{-1}$	& 9.13$\cdot10^{-2}$	& 9.13$\cdot10^{-2}$	& 1.00 			\\
  \hline
  2003cg	& 1		& 6.98$\cdot10^{-2}$	& 6.48$\cdot10^{-2}$	& 6.51$\cdot10^{-2}$	& 6.07$\cdot10^{-2}$	& 1.07			\\
  \hline
  2004W		& 1		& 3.53			& 3.21			& 1.03$\cdot10^{-1}$	& 1.30$\cdot10^{-1}$	& 7.92$\cdot10^{-1}$	\\
  \hline
  2006X		& 0		& 6.98$\cdot10^{-1}$	& 6.89$\cdot10^{-1}$	& 4.98$\cdot10^{-1}$	& 5.02$\cdot10^{-1}$	& 9.92$\cdot10^{-1}$ 	\\	\\
  \hline
  2006dd	& 6		& 6.68			& 5.37			& 1.55$\cdot10^{-1}$	& 1.55$\cdot10^{-1}$	& 1.00 			\\
  \hline
  2006mr	& 44		& 23.7			& 22.4			& 5.87$\cdot10^{-5}$	& 1.83$\cdot10^{-5}$	& 3.21			\\
  \hline
  2007gi	& 0		& 3.21$\cdot10^{-1}$	& 2.27$\cdot10^{-1}$	& 7.26$\cdot10^{-1}$	& 7.97$\cdot10^{-1}$	& 9.10$\cdot10^{-1}$ 	\\
  \hline
  2007sr	& 4		& 6.29			& 3.37			& 1.21$\cdot10^{-1}$	& 1.85$\cdot10^{-1}$	& 6.54$\cdot10^{-1}$	\\
  \hline
  2008fp	& 1		& 1.81$\cdot10^{-1}$	& 1.30$\cdot10^{-1}$	& 1.51$\cdot10^{-1}$	& 1.14$\cdot10^{-1}$	& 1.33			\\
  \hline
  2011fe	& 4		& 41.7			& 5.86			& 9.66$\cdot10^{-14}$	& 1.40$\cdot10^{-1}$	& 6.89$\cdot10^{-13}$	\\
  \hline
  2011iv 	& 243		& 198			& 197			& 1.77$\cdot10^{-4}$	& 1.28$\cdot10^{-4}$	& 1.38			\\
  \hline
  2012cu 	& 1		& 4.52$\cdot10^{-1}$	& 1.94$\cdot10^{-1}$	& 2.88$\cdot10^{-1}$	& 1.60$\cdot10^{-1}$	& 1.80 			\\
  \hline
  2012fr	& 5		& 7.97			& 3.94			& 9.26$\cdot10^{-2}$	& 1.54$\cdot10^{-1}$	& 6.01$\cdot10^{-1}$ 	\\
  \hline
\end{tabular} \label{Table:Compare_hypotheses}
\end{minipage}
\end{table*}

In general, the currently-used assumptions concerning the thermonuclear processing of accreted material on the surface of a massive WD are probably too simplistic to provide a good model for the X-ray emissions from such systems. The accretion process and the interaction between a luminous X-ray source and the material being accreted is bound to be complicated. What is needed is a better understanding of the detailed physics of the burning of hydrogen- and helium-rich material on the surface of massive WDs, realistic modelling of the radiative transfer processes in the entire WD+donor system, resulting in reliable observational predictions.

\section{Conclusions} \label{Sect:Conclusion}
With the results reported in this study, and disregarding the ambiguous case of SN2007on, we currently have thirteen pre-explosion \textit{Chandra} obervations of the positions of type Ia SNe. None of them show evidence of a supersoft X-ray progenitor. However, as discussed in Section \ref{Sect:Discussion}, the upper limits of the bolometric luminosities of even the best of these observations place only weak constraints on the supersoft X-ray characteristics of the progenitor systems of the SNe in question: for SN2007sr, SN2011fe and SN2012fr we can rule out only the brightest of the suggested canonical SSS progenitors.

Our search was initially started in an attempt to solve the question of the SD or DD nature of type Ia SN progenitors, since SD were expected to be SSSs, while DD progenitors were not. However, there may be several reasons for the lack of supersoft X-ray emissions from the progenitors of type Ia SNe, besides the obvious one that they are not SSS. A number of recent studies have shown the question of whether type Ia SN progenitors are X-ray sources to be somewhat more involved than what was initially assumed, for details see Section \ref{Sect:Discussion} in this paper and the discussion in \citep{Nielsen.et.al.2012}. Also, it has been suggested that DD progenitors may also emit soft X-rays for a significant period of time during the merger leading to the SN explosion, see \citet{Yoon.et.al.2007}. However, the luminosities expected in such cases are approximately an order of magnitude lower than for a steadily-accreting SD progenitor. Also, \citet{Di.Stefano.2010.b} found that if the DD scenario is the dominant contributor to the type Ia SN rate, then a significant population of X-ray bright proto-DD systems should exist as a result of wind accretion before the second WD forms. However, a study by \citet{Nielsen.et.al.2013} involving recent population synthesis codes fails to reproduce a significant population of SSS proto-DD progenitor systems as was suggested by \citet{Di.Stefano.2010.b}, and in any case, any emission from this type of system would cease long before the SN itself.

The campaign to use archival \textit{Chandra} images to constrain the soft X-ray characteristics of type Ia SNe continues, and the number of nearby type Ia SNe for which archival images exist continues to grow. While other studies have provided upper limits to the luminosities of indivdual type Ia SN progenitors, our archival search campaign provides the only consistently executed study of the upper limits of the X-ray and bolometric luminosities of nearby ($<$ 25 Mpc) type Ia SN progenitors.

\section*{Acknowledgments} \label{Sect:Acknowledgments}

We thank the IAU Central Bureau of Astronomical Telegrams for providing a list of SNe. This research made use of data obtained from the \textit{Chandra} Data Archive and the CIAO 4.3 software provided by the \textit{Chandra} X-ray Center.

This research is supported by NWO Vidi grant 016.093.305.

Additionally, we acknowledge Gijs Roelofs for help with this project in its early stages.

\newpage
\clearpage

\label{lastpage}

\end{document}